\documentclass[%
 reprint,
 superscriptaddress,
 amsmath,amssymb,
 aps,
 pra,
]{revtex4-2}

\usepackage{graphicx}
\usepackage{dcolumn}
\usepackage{bm}
\usepackage{ulem}

\usepackage{color}
\usepackage{xcolor}
\usepackage{booktabs}
\usepackage{amsmath}
\usepackage{commath}

\begin{document}

\preprint{APS/123-QED}

\title{Differential elasticity in lineage segregation of embryonic stem cells}

\author{Christine M. Ritter}
\thanks{These authors contributed equally to this work}
\affiliation{Niels Bohr Institute, University of Copenhagen, 2100 Copenhagen, Denmark}

\author{Natascha Leijnse}
\affiliation{Niels Bohr Institute, University of Copenhagen, 2100 Copenhagen, Denmark}

\author{Younes Farhangi Barooji}
\affiliation{Niels Bohr Institute, University of Copenhagen, 2100 Copenhagen, Denmark}


\author{Joshua M. Brickman}
\affiliation{The Novo Nordisk Foundation Center for Stem Cell Medicine, reNEW, University of Copenhagen, Blegdamsvej 3, 2200 Copenhagen, Denmark}

\author{Amin Doostmohammadi}
\thanks{These authors contributed equally to this work}
\affiliation{Niels Bohr Institute, University of Copenhagen, 2100 Copenhagen, Denmark}
\affiliation{Corresponding authors: doostmohammadi@nbi.ku.dk, oddershede@nbi.ku.dk}

\author{Lene B. Oddershede}
\affiliation{Niels Bohr Institute, University of Copenhagen, 2100 Copenhagen, Denmark}
\affiliation{Corresponding authors: doostmohammadi@nbi.ku.dk, oddershede@nbi.ku.dk}



\begin{abstract}
The question of what guides lineage segregation is central to development, where cellular differentiation leads to segregated cell populations destined for specialized functions. Here, using optical tweezers measurements of mouse embryonic stem cells (mESCs), we reveal a mechanical mechanism based on differential elasticity in the second lineage segregation of the embryonic inner cell mass into epiblast (EPI) cells - that will develop into the fetus - and primitive endoderm (PrE) - which will form extraembryonic structures such as the yolk sac. Remarkably, we find that these mechanical differences already occur during priming and not just after a cell has committed to differentiation. Specifically, we show that the mESCs are highly elastic compared to any other reported cell type and that the PrE cells are significantly more elastic than EPI-primed cells. Using a model of two cell types differing only in elasticity we show that differential elasticity alone can lead to segregation between cell types, suggesting that the mechanical attributes of the cells contribute to the segregation process. 
Our findings present differential elasticity as a previously unknown mechanical contributor to the lineage segregation during the embryo morphogenesis.
\end{abstract}

\maketitle

\begin{figure}[ht]
\centering
\includegraphics[width=1\linewidth]{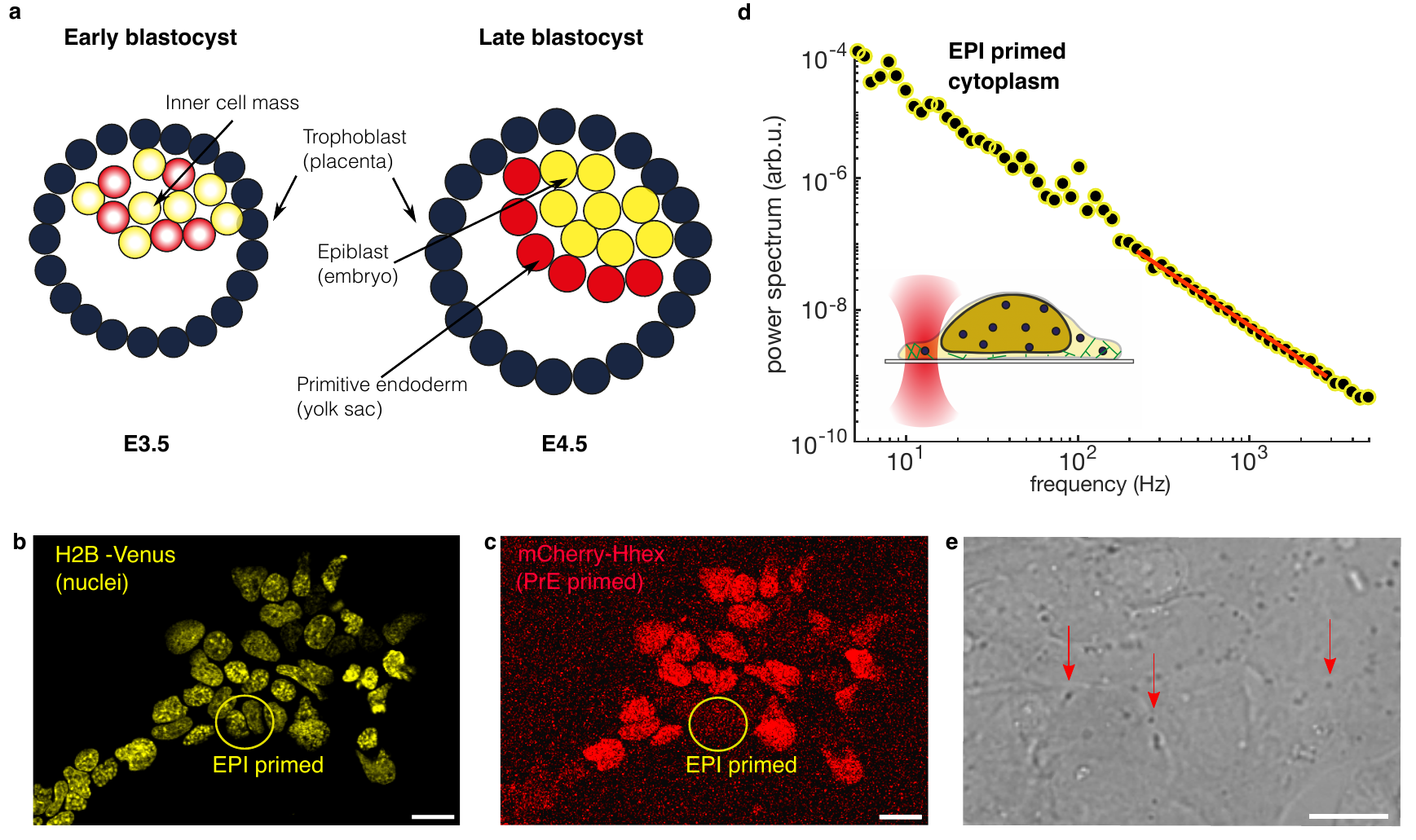}
\caption{\textbf{Autonomous segregation of the inner cell mass occurs both inside the pre-implementation mouse embryo at day E3.5 and in a culture of mESCs}. a) Schematic illustration of the segregation inside a mouse embryo between day E3.5 (left) and E4.5 (right) at which EPI-primed cells (yellow) are centrally located and PrE-primed cells(red) are located at the rim of the forming embryo. b) and c) Confocal $z$-projection of mESCs, where b) all cell nuclei are labeled with H2B-Venus (yellow) while c) the PrE-primed cells express an additional mCherry-Hhex fluorescent reporter (red). Examples of EPI-primed cells which do not express the mCherry-Hhex fluorescent reporter are marked by yellow circles in b) and c). d) Positional power spectrum of a granule located in the cytoplasm of an EPI-primed cell. The red line signifies the fitting range, 200-3000~Hz. Bottom inset: Schematic illustration of optical trapping (red: trapping laser) of a lipid granule (blue) inside the cytoplasm of an mESC (yellow). e) mESCs visualized by bright field microscopy, lipid granules are visible throughout the cells (see red arrows for examples). Scale bars are 20~$\mu$m.}
\label{fig1}
\end{figure}
Embryonic development is a paradigm of biophysical self-organization~\cite{Takaoka2012,Wennekamp2013}. The fertilized egg makes the early mammalian embryo or blastocyst as a result of several rounds of cell divisions coupled to two successive lineage choices. First, embryonic blastomeres are segregated into an inner cell mass that is enclosed by an epithelium of trophectoderm cells. The trophectoderm will go on to form the placenta, while the inner cell mass will give rise to the remaining support structures and the embryo proper. The second lineage segregation occurs when the inner cell mass differentiates into two distinct populations: epiblast (EPI) that will develop into the fetus, and primitive endoderm (PrE) that will form extraembryonic structures such as the yolk sac~\cite{Paulkin2012,Chazaud2006,Plusa2008} (Fig.~\ref{fig1}a). Recent studies have uncovered the mechanism of the first lineage segregation into trophectoderm and the inner cell mass based on the coordination between differential contractility and difference in the basal-apical polarity (pointing towards the outside of the embryo)~\cite{Maitre2016}, indicating that mechanical features of early embryos play a defining role in the first lineage decision-making~\cite{Nissen2017,Chowdhury2010,Vining2017}. However, the role of potential mechanical mechanisms for the second lineage divergence into EPI and PrE remains elusive~\cite{Filimonow2019}.
The segregation of the inner cell mass into EPI and PrE is thought to be driven by the interaction of FGF signaling with the gene regulatory networks responsible for PrE and EPI identity. The propagation of this signal via a feedback circuit leads to the induction of progenitor cells for both lineages in a salt and pepper pattern that resolves into two separate domains based on differential adhesion~\cite{Chazaud2006,Plusa2008}.
While the heterogeneity of the inner cell mass is now accepted, the differential adhesion mechanism remains controversial. In particular, recent experiments have shown no evidence of any difference between the main determinants of adhesion between PrE and EPI cells~\cite{Filimonow2019}. Similar conclusions have been drawn for epithelial cell layers, where experiments and theory refuted cell segregation based on the differential adhesion between epithelial cells~\cite{balasubramaniam2021investigating}. Additionally, differential line tension and differential affinity mechanisms have been shown to be insufficient to govern cell sorting in the inner cell mass~\cite{yanagida2022cell}. Here, we use \textit{in vitro} experiments exploiting PrE priming in mouse ESCs alongside with cell-based modeling, to show that differences in mechanical properties of the cells within the inner cell mass and in particular their distinct elastic behaviors - rather than differences in adhesion and interaction properties - could lead to the autonomous sorting of cells into PrE and EPI sub-populations.\\ 

\noindent {\bf Endodermal primed (PrE) cells are significantly more elastic than epiblast (EPI) cells}. Na\"{i}ve embryonic stem cells are {\it in vitro} cell lines derived from the {\it ex vivo} expansion of the inner cell mass. They are heterogeneous and can be shown to exhibit reversible lineage priming from an epiblast-like cell to a population biased for PrE differentiation~\cite{Riveiro2020}. We used mouse embryonic stem cells (mESC) as a model system~\cite{Paulkin2012,Huch2015}  carrying a fluorescent mCherry-Hhex reporter alongside a constitutively expressed nuclear label, H2B-Venus~\cite{Illingworth2016}. Cells exhibiting mCherry fluorescence are primed for PrE and all cells in the culture express nuclear Venus. The degree of mCherry fluorescence within a cell can be used to determine its priming  (Fig.~\ref{fig1}b,c)~\cite{Canham2010,Morgani2013,Huch2015}. Using these cells, we compared elastic properties of mCherry high and low mESCs. Experiments are performed at E3.5, which corresponds to the ``early wave'' of lineage expression and precedes the full lineage exclusive expression and positional ordering that occur at later stages~\cite{Plusa2008,mohammed2017single} (Fig.~\ref{fig1}a). 

\begin{figure*}[ht]
\centering
\includegraphics[width=0.8\linewidth]{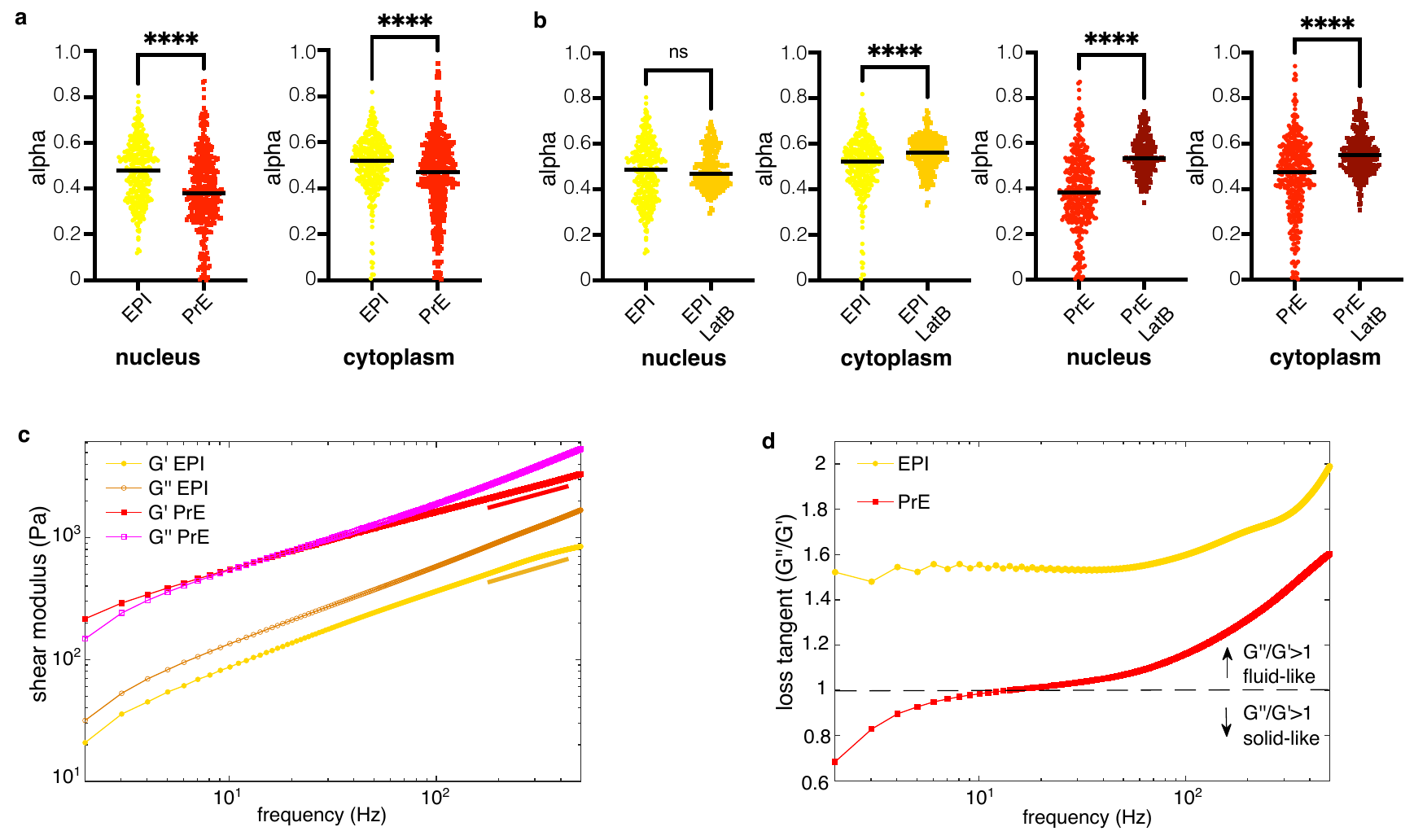}
\caption{\textbf{PrE-primed mESCs are more elastic than the EPI-primed mESCs.} a) Scatter plots of scaling exponents, $\alpha$, characterizing the sub-diffusive motion of granules inside mESCs for PrE- and EPI-primed cells inside the nuclei (left) and cytoplasm (right) show significantly lower $\alpha$ values and thus higher elasticity for PrE-primed mESCs. b) Inhibiting actin polymerization decreases mESC elasticity. Scatter plots of scaling exponents, $\alpha$, for control cells vs. cells treated with Latrunculin B (LatB) for EPI-primed cells in nucleus (1st plot) and cytoplasm (2nd plot) and in PrE-primed cells in nucleus (3rd plot) and cytoplasm (4th plot). c) Complex shear moduli, $G'$ and $G''$, calculated for EPI- and PrE-primed mESCs averaged over 10 different granules per experiment. d) Loss tangent averaged over 10 different granules per experiment. The scaling behavior of $G'(f)$ and $G''(f)$ comply well with those obtained from power spectral analysis in comparable frequency intervals. Both the absolute value of $G'(f)$ and the calculated loss tangent show that PrE-primed cells are significantly more elastic than EPI-primed cells. Black lines in scatter plots in a) and b) denote the medians. Mean, standard deviations, and $p$-values for data in a) and b) are summarized in Supplementary Tables S1 and S2.}
\label{fig2}
\end{figure*}

We used optical tweezers to trace the thermal fluctuations of nano-scopic tracer particles~\cite{pesce2020optical,marago2013optical}, in the form of endogenously occurring lipid granules (Fig.~\ref{fig1}d,e), to determine the local elastic properties within individual cells~\cite{Leijnse2012,tolic2004}. In the high-frequency regime, where the tracer does not feel the restoring force from the optical trap, the power spectrum of a diffusive tracer particle in a medium follows a power law~\cite{ghosh2016} $P(f)\propto f^{-(1+\alpha)}$, where the scaling exponent $\alpha$ carries information about the elastic properties of the medium~\cite{Schnurr1997,Selhuber2009,Ekpenyoung2012,favre2017optical,sugden2017endoglin,hurst2021intracellular}. The closer the scaling exponent $\alpha$ is to 1, the more viscous the tracer’s environment, the closer $\alpha$ is to zero, the more elastic the environment. In our study, the fitting interval for obtaining $\alpha$ (200-3000~Hz) was chosen such that it is well above 100~Hz, the frequency below which non-equilibrium processes typically occur in biological samples, and as such within this fitting interval we are probing the material properties inside of individual cells~\cite{Wessel2015,Gallet2009}. 
The scaling exponents measured for all mESCs were significantly lower than values reported for any other cell type, showing that, overall mESCs are highly elastic (see Fig. S1 for a comparison of the scaling exponent to epithelial, endothelial, fibroblast, and yeast cells). Interestingly, the average value of the scaling exponents in all PrE-primed cells $\alpha$=$0.42\pm0.17$ (mean $\pm$ s.d.) was significantly smaller than the value from EPI-primed cells $\alpha$=0.49$\pm$0.13 (mean $\pm$ s.d.) (Fig.~\ref{fig2}a and Fig.~S2), indicating that the PrE-primed cells are more elastic than the EPI-primed population at this early stage of embryonic development. This is important because our measurements show that these observed significant differences in the elasticity are already occurring during priming and not just after a cell has committed to differentiation.

To further validate the results obtained from the scaling exponent analyses
we next calculated the complex shear modulus $G=G'+iG''$ of the lineage primed embryonic stem cells (Fig.~\ref{fig2}c and Fig.~S3) as the defining characteristics of viscoelastic materials (see Supplementary Material for details of the measurements)~\cite{Mason1995,Borries2020}. The storage modulus $G'$, characterizes the energy stored in the material during a cycle of deformation, and therefore is a measure of the elastic component of the material behavior, while the loss modulus $G''$, characterizes the average dissipation of the energy and is a measure of the fluidity of the material~\cite{bird1987}. Therefore, in order to assess the relative fluid-like and solid-like behavior of the cells we measured the loss tangent $G''/G'$, the ratio of the loss to storage modulus for EPI- and PrE-primed cells (Fig.~\ref{fig2}d). Consistent with the results of scaling exponent measurements from power spectral analyses, 
a significantly higher elasticity was observed for PrE-primed cells compared to EPI-primed ones.
Together, our analyses of the complex shear modulus confirm the results of power spectral analyses, demonstrating that (i) the mESCs are highly elastic compared to other cell types and (ii) there is a significant difference between the elasticity of PrE- and EPI-primed cells, the behavior of which resemble the inner cell mass at the early stages of the embryo development~\cite{Chazaud2006}.\\

\noindent {\bf Both nucleus and cytoplasm of primed PrEs are significantly more elastic than in primed EPIs.} Next, in order to further assess the elasticity difference between the two sub-populations within lineage primed mESCs, we compared the internal structure of the two sub-populations. Fluorescent labeling of the nucleus (using a H2B-Venus expression construct) allowed for its independent visualization. 
In addition, in our optical tweezers experiments, the presence of lipid granules both in the nucleus and in the cytoplasm, allowed for spatially resolved measurement of the elasticity locally within the cells (see Supplementary Material for details). For both cell types, the scaling exponents characterizing the movement of tracers in the nuclei were found to be significantly lower than in the cytoplasm (Fig.~\ref{fig2}a), demonstrating that for all probed cell types, the nucleus was more elastic than the cytoplasm (see values of $\alpha$ in Supporting Table S1 and results of $t$-tests in Supporting Table S2). This observation is in line with a previous study of articular chondrocytes using micro-pipette aspiration, which showed that the nuclei were significantly stiffer than the cytoplasm~\cite{Guilak2000}. 
Comparing the spatially resolved scaling exponents between the nuclei of PrE- and EPI-primed cells, we found that both the nuclei and cytoplasm of the primitive PrE-primed cells were significantly more elastic than the EPI-primed cells. Together, these results show that the observed differential elasticity between the mESCs primed for EPI and the PrE is consistent for comparison of cytoplasm, nuclei, and of entire cells.\\

\noindent {\bf Higher levels of polymerized actin in PrE-primed cells are linked to their higher elasticity compared to EPI-primed cells.} The difference in elasticity of EPI- and PrE-primed cells may be attributed to the difference in structure and composition of their subcellular elements. Although actin is predominantly present at the periphery of the cells~\cite{Citters2006}, in ESCs it is also present in the nucleus where it has been shown to associate with lamin A and chromatin remodeling complexes~\cite{Chalut2012}. We hypothesized that a higher level of polymerized actin filaments in the PrE-cells may produce the difference in the elasticity between EPI- and PrE-primed cells. To test this hypothesis we treated cells with the actin disruptive drug, Latrunculin B (LatB) (see Supplementary Material for the details of drug treatments). 
As expected, for all investigated categories (nucleus or cytoplasm, EPI- or PrE- primed), the cells became less elastic (showed a higher value of the scaling exponent $\alpha$) upon LatB treatment, consistent with removal of actin filaments (Table S1, Table S2). Importantly, however, while actin disruption led to a small decrease in the elasticity of the EPI-primed cells, PrE-primed cells became significantly less elastic (Fig.~\ref{fig2}b, Fig.~S4, and Table S1), indicating that higher levels of actin filaments in these PrE-primed sub-populations governs their higher elasticity. Together, these results suggest that the difference in internal molecular composition and structure of the primed cells lead to the difference in their elasticity, showing that changes in internal composition and structure already occur during priming and not just after a cell has committed to differentiation.\\
\begin{figure}[b]
\centering
\includegraphics[width=1\linewidth]{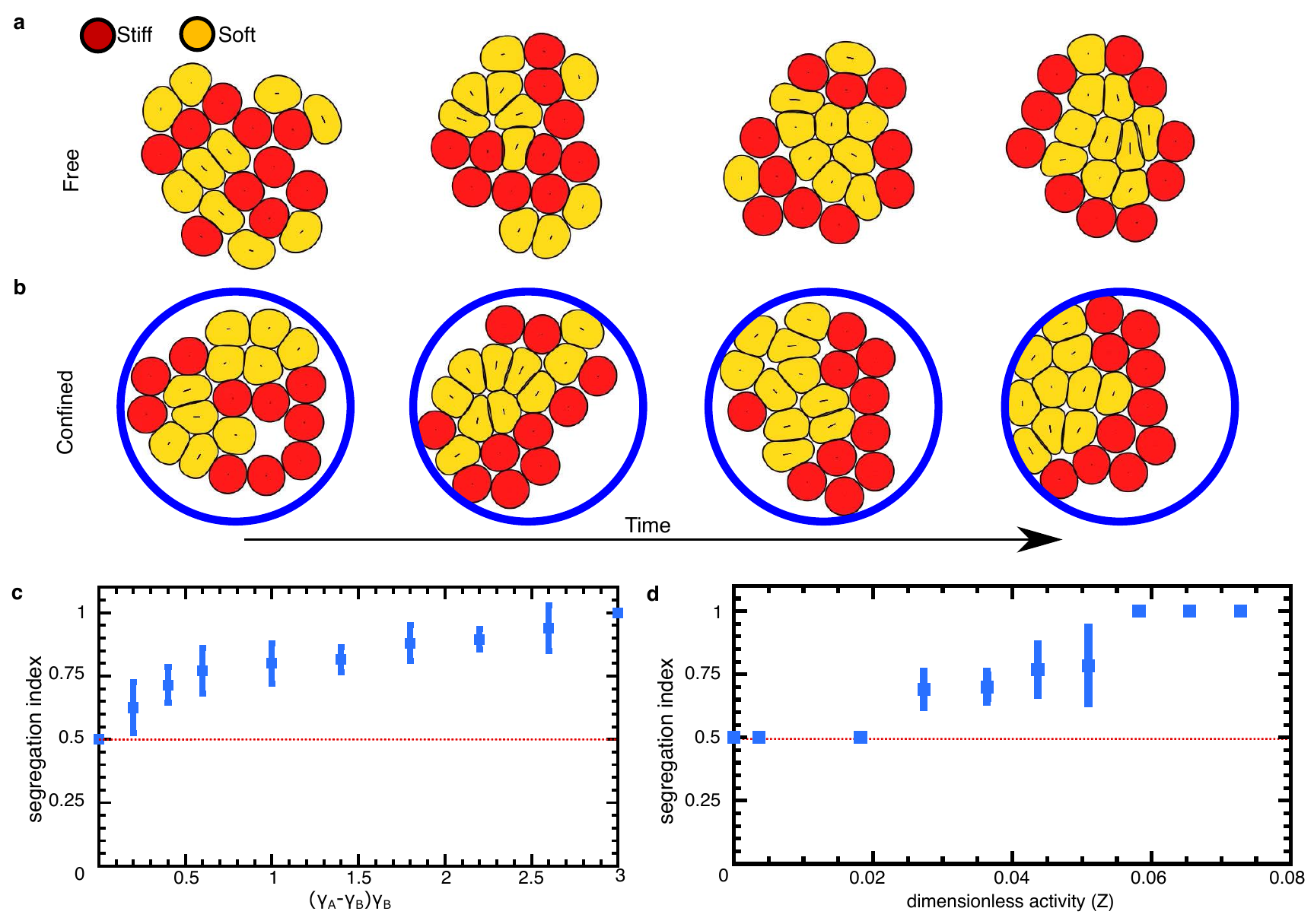}
\caption{\textbf{Differential elasticity is an active mechanism for cell segregation.} a,b) Time series snapshots of the segregation of a 20-cell aggregate of soft (yellow) and stiff (red) cells simulated a) without any confining boundaries, and b) in a circular confinement. c) Cell segregation as a function of the normalized elasticity difference $(\gamma_A-\gamma_B)/\gamma_B$, where $\gamma_A$ denotes the elasticity of the stiff cells and $\gamma_B$ is the elasticity of the soft cells. e) Cell segregation as a function of the strength of active stresses generated by the cells. The segregation index $SI=\langle N_A/(N_A+N_B)\rangle$ is defined such that $SI=1/2$ when the outer layer is a uniform mixture of stiff and soft cells (indicated by the dashed red line), $SI=1$ when the outer layer is occupied solely by stiff cells, and $SI=-1$ if only soft cells occupy the outer layer. Error bars indicate the variance computed from averaging over 5 different simulations with random initial positioning of the cells.}
\label{fig3}
\end{figure}

\noindent {\bf Minimal physical model shows elasticity difference between PrE- and EPI-primed cells is sufficient for the observed cell segregation.} When EPI and PrE primed mESCs are injected into host embryos they colonize their respective lineage and when mixed in aggregation culture, they segregate, recapitulating the differential localization of PrE and EPI \textit{in vivo}~\cite{Shabazi2019}. Could the difference in elasticity alone be sufficient for cell segregation? In order to answer this question, we developed a cell-based model of the inner cell mass composed of two cell types that only differ in their elastic properties. We build on a phase-field model~\cite{Mueller2019,balasubramaniam2021investigating} that captures the mechanical properties of the cells, including elasticity, deformability and contractility, at the single cell level and allows for accurate representation of boundaries between the cells~\cite{Palmieri2015,wenzel2021multiphase,jain2022impact} (see Supplementary Material for detailed description of the model). Importantly, this model allows us to vary the elasticity, while explicitly keeping the same interaction forces, i.e., adhesion and repulsion forces, between the cells. In brief, each cell is represented by a phase-field parameter $\Phi$ that distinguishes the interior $\Phi_i=1$, and the exterior $\Phi_i=0$, of an individual cell. The dynamics of each phase-field evolves according to the Cahn-Hilliard equation:
\begin{align}
    \partial_t\Phi_i+\bm {v_i}\cdot\nabla\Phi_i=-\frac{\delta \mathcal{F}}{\delta \Phi_i}, \label{eqn:1}
\end{align}
 where the right-hand-side describes the relaxational dynamics towards the minimum of a free energy $\mathcal{F}=\mathcal{F}_{\text{cell}}+\mathcal{F}_{\text{int}}$, that is composed of two parts: $\mathcal{F}_{\text{cell}}$ determines the mechanical properties of the cell including its elasticity, compressibility and deformability; $\mathcal{F}_{\text{int}}$ sets the strength of adhesive and repulsive interactions between the cells. In addition to these equilibrium contributions, the advective term $\bm {v_i}\cdot\nabla\Phi_i$, accounts for the activity of each cell that is determined from the force balance $\xi \bm {v_i} =\bm{F}_i^{\text{int}}+\bm{F}_i^{\text{active}}$, between the active forces generated by the cells $\bm{F}_i^{\text{active}}$ that include contractility~\cite{prost2015active,paluch2006dynamic,salbreux2012actin,chugh2018actin}, substrate friction $\xi$, and the cell-cell interaction forces $\bm{F}_i^{\text{int}}$ (see Supplementary Material for the definitions of individual terms). 
 
 In order to isolate the effect of elasticity, all the physical parameters related to the cells contractility, and cell-cell interactions, are imposed as identical for both cell types. Simulations are initialized with a mixture of two cell types with different elasticity positioned randomly within the domain (Fig.~\ref{fig3}a). In agreement with the experiments, the cells autonomously self-segregate with time. More importantly, the cells organize themselves into a batch of less elastic cells that are ensnared with stiffer cells at the outer ring (Fig.~\ref{fig3}a,b). This form of the cell segregation is 
 consistent with the well-established \textit{in vivo} behavior in the blastocyst where the PrE-lineage forms a layer separating the epiblast from the blastocoel~\cite{Chazaud2006,Shabazi2019}. 
 
 To quantify the cell segregation we defined the segregation index, $SI=\langle N_A/N\rangle$, as the fraction of the outer cell layer that is occupied with elastic cells, where $\langle N_A \rangle$ is the time-averaged number of the elastic cells in the outer layer and $N=N_A+N_B$ is the total number of the cells in the outer layer. Therefore, for a perfect segregation with the more elastic cells in the outer layer $SI=1$, while $SI=1/2$ for a perfectly mixed population, and $SI=0$ corresponds to a perfect segregation in which the less elastic cells form the outer layer. The measurements of the segregation index for varying elasticity difference between the cells showed that an elasticity difference as small as 10~\% is enough to trigger the cell segregation (Fig.~\ref{fig3}c). We further checked that the observed cell segregation is independent of the type of the boundary and the size of the system (Fig.~\ref{fig3}b, Fig.~S5).
 
In addition, it is possible to pinpoint the key mechanism of the cell sorting in the simulations. Interestingly, we found that cell sorting based on the difference in the elasticity is only achieved in the presence of the active stress generation by the cells (Fig.~\ref{fig3}d). The existence of such active stresses due to the contractile tension at the cell cortex and the myosin activity is well documented~\cite{paluch2006dynamic,salbreux2012actin,chugh2018actin,yanagida2022cell}. In our simulations we find that as the cells continuously push and pull on their neighbors by generating active stresses, the less elastic cells are able to deform more, exchange neighbors at a higher rate (Fig.~S6), and therefore form an inner cell aggregate that shows a fluid-like behavior, while the more elastic cells reside at the outer layer and do not do a considerable neighbor exchange (Fig. S5), thus exhibiting a more solid-like behavior. Therefore, our modeling results suggest that the cell segregation based on differential elasticity of the cells is an inherently out of equilibrium phenomenon, which cannot be explained by purely thermodynamic processes.

Together our results show that the observed difference in material properties, with the PrE primed mESCs being significantly more elastic than those cells primed for EPI differentiation, could account for the capacity of these cells to segregate when reintroduced into blastocysts~\cite{Canham2010}.   As this ESC heterogeneity reflect the transcriptional and signaling events occurring in the inner cell mass of the blastocyst, this suggests that differential elasticity could function in conjunction with other mechanisms such as differential adhesion~\cite{Chazaud2006,Plusa2008}, to govern the second lineage segregation in mammalian embryogenesis. This differential elasticity mechanism reinforces a series of recent findings that showed the importance of mechanical mechanisms such as fluid-to-solid jamming transition in determining the body axis elongation in vertebrates~\cite{Mongera2018}, 
and differential contractility in governing the first lineage segregation in mouse preimplantation development~\cite{Maitre2016}. Moreover, the essential role of difference in material properties~\cite{tanaka2022viscoelastic} combined with active stress generation by the cells indicates that the non-equilibrium physics of active materials might have played a crucial role in the context of autonomous cell sorting during development.\\

\section*{Acknowledgments}
We thank William B. Hamilton and Romain Mueller for discussions and help during the early stages of the work. The work was financially supported by the Danish National Research Council through the StemPhys Center of Excellence, grant nr DNRF116. A. D. acknowledges funding from the Novo Nordisk Foundation (grant No. NNF18SA0035142 and NERD grant No. NNF21OC0068687), Villum Fonden Grant no. 29476, and the European Union via the ERC-Starting Grant PhysCoMeT. 

\bibliography{apssamp}

\end{document}